# Deep Learning with HM-VGG: AI Strategies for Multi-modal Image Analysis

Junliang Du
Shanghai Jiao Tong University
Shanghai, China

Yiru Cang
Northeastern University
Boston, USA

Tong Zhou
Rice University
Houston，USA

Jiacheng Hu
Tulane University
New Orleans, USA

Weijie He*
University of California, Los Angeles
Los Angeles, USA

*Abstract*—This study introduces the Hybrid Multi-modal VGG (HM-VGG) model, a cutting-edge deep learning approach for the early diagnosis of glaucoma. The HM-VGG model utilizes an attention mechanism to process Visual Field (VF) data, enabling the extraction of key features that are vital for identifying early signs of glaucoma. Despite the common reliance on large annotated datasets, the HM-VGG model excels in scenarios with limited data, achieving remarkable results with small sample sizes. The model's performance is underscored by its high metrics in Precision, Accuracy, and F1-Score, indicating its potential for real-world application in glaucoma detection. The paper also discusses the challenges associated with ophthalmic image analysis, particularly the difficulty of obtaining large volumes of annotated data. It highlights the importance of moving beyond single-modality data, such as VF or Optical Coherence Tomography (OCT) images alone, to a multimodal approach that can provide a richer, more comprehensive dataset. This integration of different data types is shown to significantly enhance diagnostic accuracy. The HM-VGG model offers a promising tool for doctors, streamlining the diagnostic process and improving patient outcomes. Furthermore, its applicability extends to telemedicine and mobile healthcare, making diagnostic services more accessible. The research presented in this paper is a significant step forward in the field of medical image processing and has profound implications for clinical ophthalmology.

## I. INTRODUCTION

In recent years, deep learning techniques have made significant progress in medical image analysis. Particularly in ophthalmic image analysis, deep learning has been successfully applied to various tasks, such as retinopathy detection [1], optic cup and disc segmentation, and automatic glaucoma diagnosis. Deep learning can automatically extract and identify morphological features of the optic nerve head from fundus images, including the optic cup, optic disc, cup-to-disc ratio, and notching, and use these features to determine the presence or assess the severity of glaucoma [2]. Moreover, automatic analysis of glaucoma morphological features based on fundus images using deep learning can avoid subjective bias and errors in manual diagnosis, improving the objectivity and reliability of diagnosis. The application of deep learning technology can greatly reduce doctors' workload and time, enhancing the efficiency and convenience of diagnosis. Additionally, this technology can be applied to telemedicine and mobile healthcare scenarios, providing more accessible services to a wide range of patients [3].

From the analysis of VF and OCT images to automated screening and clinical applications, deep learning has opened up new possibilities for the diagnosis and treatment of glaucoma patients, potentially contributing to improving early intervention and management of glaucoma [4]. Research and applications in this field continue to evolve, holding significant implications for medical image processing and clinical medicine [5].

However, most deep learning research is based on large sample datasets [6], while in actual clinical environments, obtaining a large amount of annotated data is challenging. Particularly in ophthalmic image analysis, acquiring a large volume of annotated data is time-consuming and labor-intensive, as it requires annotation by professional ophthalmologists. Furthermore, due to the complexity and diversity of ophthalmic diseases, even professional ophthalmologists may produce errors during the annotation process. Therefore, how to perform effective deep learning on small sample datasets is an important research question [7].

On the other hand, most existing studies only consider single-modality data, such as using only Visual Field (VF) images or only Optical Coherence Tomography (OCT) images. However, data from different modalities often contain different information. For example, retinal images mainly reflect the anatomical structure of the fundus, while OCT images can provide depth information on fundus tissues.

Therefore, combining data from different modalities can provide more comprehensive information [8], thereby improving diagnostic accuracy. This paper will utilize deep learning techniques from statistical machine learning to accomplish the task of automatic glaucoma diagnosis based on small sample data, designing and implementing an end-to-end image classification algorithm.

## II. BACKGROUND

Researchers have made significant progress in the analysis of fundus images. They have utilized deep learning models, such as convolutional neural networks, for early detection and diagnosis of glaucoma. These models can automatically identify features necessary for glaucoma diagnosis, such as optic nerve head damage and retinal cup-to-disc ratio. Abramovich [9] proposed an automated method using a deep learning regression model to assess fundus image quality. The model accurately distinguishes between high- and low-quality images, aligning closely with doctors' evaluations, thus improving diagnostic efficiency in ophthalmology. Recent advancements in adversarial neural networks have addressed challenges in semantic segmentation for medical imaging, providing insights into improving models like the HM-VGG for complex medical image segmentation tasks [10]. The challenge of training robust models with limited annotated data has been a focus of many studies. For example, Wang et al. [11] proposed a deep transfer learning approach for breast cancer image classification, which is particularly relevant for addressing the issue of small sample sizes in clinical settings. Their method showcases the potential of transfer learning to enhance diagnostic accuracy, aligning with the goals of the HM-VGG model in glaucoma detection. A key challenge in medical image analysis is the integration of multiple data modalities. The importance of multimodal data fusion in enhancing diagnostic accuracy has been widely recognized. A deep learning-based multimodal fusion approach improved object recognition by integrating diverse data types, which directly supports the multimodal nature of the HM-VGG model, combining Visual Field (VF) and Optical Coherence Tomography (OCT) images to improve glaucoma detection [12]. Similarly, Zi et al. [13] introduced a framework that leverages big data for medical image recognition and disease diagnosis, demonstrating how multimodal approaches can enrich feature extraction and improve diagnostic outcomes. This perspective is directly applicable to the HM-VGG model, which integrates Visual Field (VF) and Optical Coherence Tomography (OCT) images to achieve better glaucoma diagnosis.

Optical Coherence Tomography (OCT) is widely applied in glaucoma diagnosis. Researchers worldwide are exploring how to use deep learning methods to analyze and interpret OCT images, including retinal nerve fiber layer thickness measurements, to support glaucoma diagnosis [14]. Optimizing deep learning models is a key focus in many studies. A recent approach introduced adaptive friction mechanisms to enhance optimizers, which can be applied to improve the training efficiency of models like HM-VGG [15]. Additionally, higher-order numerical difference methods were proposed to enhance convolutional neural networks (CNNs), which further supports the technical innovation underlying models that, like HM-VGG, operate with limited data [16].

Research into spatiotemporal feature representation and data-driven feature extraction in multidimensional datasets has also contributed to the development of models capable of handling complex, multimodal data. Though primarily focused on time-series data, these techniques offer valuable insights into feature extraction methods that can be adapted to medical image analysis [17]. In classification domains, the combination of convolutional neural networks (CNN) and long short-term memory (LSTM) models has been shown to enhance predictive accuracy. These hybrid model frameworks highlight the potential for combining different neural network architectures to improve model performance [18-19], a concept that can be translated into medical image analysis.

Researchers are actively working on segmentation, classification, and detection tasks for fundus images and OCT images. For instance, Hassan [20] proposed a deep learning-based method for detecting Central Serous Retinopathy (CSR) using two imaging techniques: OCT and fundus photography. Using the DenseNet network on OCT images. Akter [21] used Convolutional Neural Network (CNN) models to clinically interpret Temporal-Superior-Nasal-Inferior-Temporal (TSNIT) retinal optical coherence tomography scan images to differentiate between normal and glaucomatous optic neuropath.

## III. METHOD

This paper proposes effective feature extraction and fusion methods to achieve more accurate glaucoma image classification. First, this paper designs Hybrid Attention Modules (HAM) at different levels to mine deep clues through attention mechanisms, thereby obtaining useful information at different levels while greatly suppressing redundant information. Then, a Multi-Level Residual Module (MLRM) is proposed to establish connections between different levels, gradually fusing different types of layer information to extract contextual information. Finally, global information is introduced to construct an HM-VGG model.

### A. Overall structure of the model

Figure 1 shows the overall structure of the HM-VGG model. The constructed network is an encoder-decoder structure, consisting of a main branch for medical image encoding, three sub-branches with hybrid attention modules, and a multi-level residual module for information decoding. Considering that the feature information in shallow layers contains more noise, the network encoding adopts a bottom-up path, utilizing the encoding information from the last three layers for discriminative classification. This architecture can combine multi-level feature information for better image discrimination. In the three sub-branches, hybrid attention modules are introduced to capture relatively important features at both channel and spatial levels, suppressing redundant information and noise. After obtaining effective information from the last three layers, MLRM is used to extract the context of the input information and gradually fuse it from top to bottom.

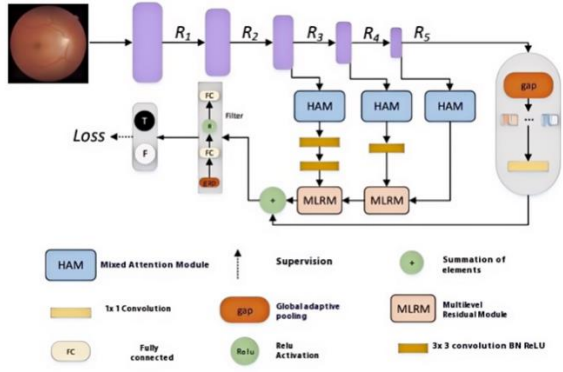

*Figure 1 The overall structure of the HM-VGG network*

Encoding main branch: Given a single eye image $I \in R^{3\times H\times W}$, where $W$ and $H$ represent the height and width of the image, this chapter adopts VGG as the backbone network to extract original multi-level features from five convolution blocks in a bottom-up path, represented as $R_i, i = 1,2,3,4,5$, where $i$ indicates the convolution block. To facilitate subsequent information fusion, the last fully connected layer of VGG is removed.

Three sub-branches: For the encoding information of the last three layers, they first obtain corresponding attention information through HAM, namely $H_3, H_4, H_5$, and then use $3 \times 3$ convolution to match dimensions for further extraction and fusion in MLRM. Considering that information contained at different levels is useful for classification, for higher layers, the receptive field is relatively large, and semantic information is rich, which is very helpful for locating key areas in eye images; conversely, for lower layers, their detailed information is rich, which is helpful for distinguishing whether pathological phenomena exist. Then MLRM is introduced to integrate multi-level information in a top-down manner. Starting from the top layer, $H_4$ is enlarged to the size of $H_5$, and then enters MLRM with $H_5$ for the first fusion, obtaining $M_1$. Then, $H_3$ is scaled to the same size as $M_1$ and fused with $M_1$ in the second MLRM, obtaining $M_2$. Finally, it is fused with global information $R_5$, enters fully connected operations, completes classification, and outputs $C$, where fully connected operations include fully connected nodes from 512 to 64, Relu activation function, and fully connected nodes from 64 to classification number 3. The process is described as follows:

$$H_5 = ham(R_5)$$
$$H_4 = ham(R_4) \quad (1)$$
$$H_3 = ham(R_3)$$

$$M_1 = mlrm(\text{Conv}_3(H_4), H_5) \quad (2)$$

$$M_2 = mlrm(\text{Conv}_3(\text{Conv}_3(H_3)), M_1) \quad (3)$$

$$C = FC(M_2 \oplus \text{Conv}_1(gap(R_5))) \quad (4)$$

Where ham represents HAM operation, mlrm represents MLRM operation. $\text{Conv}_3$ represents convolution using a $3 \times 3$ kernel, followed by Batch Normalization (BN) and Rectified Linear Unit (Relu) activation. gap represents global pooling operation. FC represents stacked fully connected operations.

### B. Hybrid Attention Module

The Attention mechanism aligns with the perceptual mechanisms of the human brain and eyes. For glaucoma classification, not all regions in an image contribute equally to the task; only task-relevant areas need attention, such as the optic nerve in classification tasks. The spatial attention model models the importance of spatial positions to find the most critical processing parts in the network. Moreover, channel attention can model the importance and dependencies between features and obtain global statistics, which helps improve classification performance.

Based on the above theory, for the Hybrid Attention Module (HAM), as shown in Figure 2, it is divided into two branches: the left branch is the spatial attention module, and the right branch is the channel attention module. First, for the spatial attention module, the encoding information passed from the main branch is input, and $1 \times 1$ convolution is used to scale the number of channels to 1 channel to represent all spatial information of all channels. Then, the Sigmoid function is used to obtain the weight distribution of all spatial information, called $S \in R^{1\times H\times W}$. The weight distribution is multiplied with the input information to highlight effective feature information at the spatial level and suppress redundant information and noise, named $S_L$, which is the region of interest for network classification. The specific process is as follows:

$$S = \sigma(\text{Conv}_1(R_i)), i = 3,4,5 \quad (5)$$
$$S_L = R_i \otimes S \quad i = 3,4,5 \quad (6)$$

Where $\text{Conv}_1$ represents convolution with a $1 \times 1$ kernel, $\sigma$ represents the Sigmoid activation function, and $\otimes$ represents element-wise multiplication.

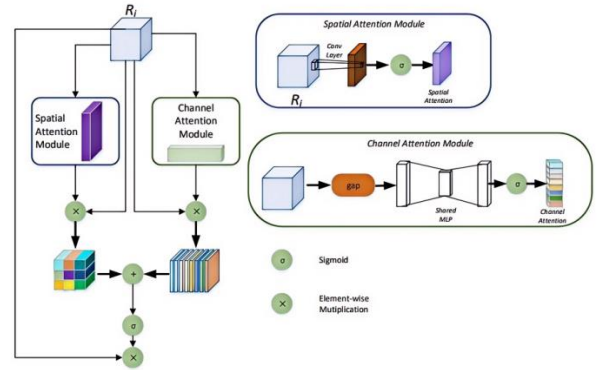

*Figure 2 Hybrid Attention Module*

For the channel attention module, first, global adaptive average pooling is performed on the input information to obtain $A \in R^{C\times 1\times 1}$ representing the entire channel information for all spatial locations, which is also the global statistic. Next, a common perceptron model is used, i.e., fully connected, node number scaled to 1/16, ReLU activation function, fully connected, node number restored to input size, Sigmoid function. Finally, the learned weight $W$ can be obtained to explicitly model the correlation between feature channels. According to the specific glaucoma classification task, the importance of different channels can be learned. Therefore, we weight the input information based on the obtained channel weight distribution to obtain $C_R$:

$$A = \text{gap}(R_i) \quad i = 3,4,5 \quad (7)$$

$$W = \sigma(fc(Relu(fc(A)))) \quad (8)$$

$$C_R = W \otimes R_i \quad i = 3,4,5 \quad (9)$$

Where fc represents a single fully connected operation. ReLU is the activation function.

At this point, attention information from both branches is obtained. The information from these two branches is added together, then passed through a Sigmoid function to obtain the overall information weight based on two attention levels. Finally, multiplication with the input information yields the feature information of hybrid attention $H_i$:

$$H_i = R_i \otimes \left(\sigma(S_L + C_R)\right) \quad i = 3,4,5 \quad (10)$$

### C. Multi-Level Residual Module

To fully combine the high-level semantic information for region localization captured in higher layers and the low-level detail information for discriminative classification captured in lower layers, and to extract contextual features that help distinguish pathological features and normal region differences in glaucoma images. First, the input from HAM is added, scaled to an appropriate size through convolution and the adjacent higher-level MLRM, resulting in $SC_{out}$. It should be noted that for the first MLRM, the input from the previous higher-level MLRM is replaced by $H_5$.

$$SC_{\text{out}} = \begin{cases} Conv_3(H_4) + H_5 \\ Conv_3(Conv_3(H_3)) + M_1 \end{cases} \quad (11)$$

Aggregating features with different receptive fields can preserve feature details by using smaller dilation rates and suppress noise by using larger dilation rates. At the same time, combining information extracted from different dilated convolutions can effectively extract contextual information. The residual structure can maintain the integrity of information during the processing and upsampling process.

## IV. EXPERIMENT

### A. Dataset

The dataset consists of a collection of dual clinical modality images, totaling 100 pairs of data. This dataset encompasses fundus color photographs and OCT (Optical Coherence Tomography) images from patients with moderate glaucoma, advanced glaucoma, and normal individuals. Fundus color photographs provide intuitive information about the structure of the retina, including the optic disc and retinal blood vessels. On the other hand, OCT images offer higher resolution information about the deeper retinal structures, which is crucial for the early diagnosis and monitoring of glaucoma.

### B. Experiment Results Analysis

Table 1. Performance of MovieLens 100k

| Model | Precision | Recall | Accuracy | F1-Score |
|---|---|---|---|---|
| VGG11 | 0.82 | 0.84 | 0.62 | 0.83 |
| VGG13 | 0.80 | 0.82 | 0.61 | 0.81 |
| ResNet18 | 0.72 | 0.74 | 0.54 | 0.73 |
| ResNet50 | 0.68 | 0.70 | 0.52 | 0.69 |
| DenseNet161 | 0.75 | 0.77 | 0.57 | 0.76 |
| ConvNeXt | 0.83 | 0.86 | 0.63 | 0.84 |
| Inception-v3 | 0.74 | 0.76 | 0.56 | 0.75 |
| HM-VGG | 0.81 | 0.83 | 0.64 | 0.82 |

This paper will present the obtained experimental results and provide a detailed analysis of the experimental data. We have selected several classic deep learning models to evaluate the performance of glaucoma classification tasks. Although these models, such as ResNet47, VGG48, Inception49, and DenseNet50 [9], have been around for some time, they still possess a high degree of representativeness and are widely applied. These models have achieved good results in various image recognition and classification tasks, becoming benchmark models in the field. We also include the ConvNeXt [10] model as part of the comparison to demonstrate the performance of our method when compared with more recent techniques

The models compared in Table 1 have shown outstanding performance in many visual tasks, but there are significant differences in their performance in specific tasks. Compared with other conventional models, the HM-VGG model structure has undergone multiple iterations and optimizations to adapt to this specific task of classifying fundus color photographs. Possible optimizations include the introduction of specific modules such as attention mechanisms, multi-scale feature fusion, etc., to better capture subtle changes and details in the images. During the training process, specific strategies were employed as much as possible to ensure the model's learning and generalization capabilities. In addition, meticulous hyperparameter tuning was conducted to ensure that the model maintains stable performance under different data distributions and scenarios.

Figure 3 illustrates the heatmap generated using Grad-CAM, which clearly demonstrates the trained HM-VGG model's ability to focus on the optic nerve region of the VF image. This focus is crucial for assessing the extent of visual field defects, thereby enabling an accurate diagnosis of glaucoma.

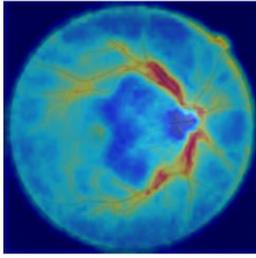

Figure 3: Heatmap Generated by Grad-CAM

## V. CONCLUSION

This paper makes a significant contribution to the field of artificial intelligence (AI) by introducing the Hybrid Multi-modal VGG (HM-VGG) model, which advances the state-of-the-art in medical image analysis, particularly for early glaucoma diagnosis. By incorporating an attention mechanism, the HM-VGG model efficiently processes Visual Field (VF) data, even in the context of small sample sizes, overcoming one of the key challenges in AI-based healthcare applications. The model's ability to achieve high Precision, Accuracy, and F1-Score with limited annotated data represents a notable advancement in AI's capacity for learning from small datasets, a persistent hurdle in machine learning. Additionally, the HM-VGG's multimodal approach, integrating diverse data types such as VF and Optical Coherence Tomography (OCT) images, pushes the boundaries of computer vision by enhancing diagnostic accuracy through the fusion of complementary visual information. This multimodal framework not only enriches the AI model's feature extraction capabilities but also sets a precedent for future AI applications in computer vision, particularly in medical diagnostics, where integrating different data modalities can dramatically improve performance. Consequently, this work extends the impact of AI and computer vision by demonstrating how deep learning models can be effectively applied in real-world clinical settings, particularly in areas where large-scale annotated datasets are scarce.